# Changes in adsorption heights upon self-assembly of bicomponent supramolecular networks

*Elizabeth Goiri,[a] Manfred Matena,[a] Afaf El-Sayed,[b] Jorge Lobo-Checa,[c] Patrizia Borghetti,[a,c] Celia Rogero,[a,c] Blanka Detlefs,[d] Julien Duvernay,[d] J. Enrique Ortega,[a,b,c], Dimas G. de Oteyza [c*]*

[a] Donostia International Physics Center, Paseo Manuel Lardizabal 4, 20018 San Sebastián, Spain

[b] Universidad del País Vasco, Dpto. Física Aplicada I, 20018 San Sebastián, Spain

[c] Centro de Física de Materiales-Materials Physics Center (CSIC/UPV-EHU), Manuel Lardizabal 5, 20018 San Sebastián, Spain

[d] European Synchrotron Radiation Facility, BP 220, 6 rue Jules Horowitz, 38043 Grenoble, France

* d_g_oteyza@ehu.es

Codeposition of two molecular species [CuPc (donor) and PFP (acceptor)] on noble metal (111) surfaces leads to the self-assembly of an ordered mixed layer with maximized donor-acceptor contact area. The main driving force behind this arrangement is assumed to be the intermolecular C-F ··· H-C hydrogen-bond interactions. Such interactions would be maximized for a coplanar molecular arrangement. However, precise measurement of molecule-substrate distances in the molecular mixture reveals significantly larger adsorption heights for PFP than for CuPc. Most surprisingly, instead of leveling to increase hydrogen bond interactions, the height difference is enhanced in mixed layers as compared to the heights found in single component CuPc and PFP layers, resulting in an overall reduced interaction with the underlying substrate. The influence of the increased height of PFP on the interface dipole is investigated through work function measurements.

INTRODUCTION

The electronic properties of metal/organic interfaces define the charge carrier injection barriers in organic optoelectronic devices, which are key parameters for their efficiency. Such properties depend largely on the detailed interfacial structure, defined by the molecular orientation with respect to the substrate, the lateral distribution and the molecule-substrate distance. The latter in particular plays a central role in the interfacial energy level alignment. This can be directly concluded from the electronic interface model put forward by Vazquez et al.,[1,2] which includes all of the commonly assumed processes determining potential changes at non-chemisorptive interfaces, such as permanent molecular dipoles, the "pillow effect" and interfacial charge transfer. The pillow effect arises from the compression, upon molecular adsorption, of the electron wavefunctions decaying into the vacuum from the metal surface. This compression modifies the surface component of the work function of the metal, and, being driven by the Pauli repulsion exerted by the adsorbates, it depends on the molecule-substrate distance. Charge transfer is determined by the induced density of interfacial states (density of states in the molecular gap, related to the shift and broadening of the molecular levels interacting with the metal) and a screening parameter. Both depend on the molecule-substrate distance as well. The most precise way to determine that distance experimentally is by means of normal incidence X-ray Standing Waves (XSW).[3,4] This technique has been applied to a number of molecules on various surfaces.[5-13] Here we report the use of XSW to determine the molecule-substrate distances in donor-acceptor molecular blends, which are not only highly relevant for many organic devices, but also have interfacial properties often differing from those of the corresponding single component layers.[14]

In this work we provide a complete structural characterization of the first monolayer of a stoichiometric 1:1 donor-acceptor mixture assembled on Ag(111) and Cu(111) substrates. The molecules used are copper phtalocyanine (CuPc, donor) and perfluoropentacene (PFP, acceptor), two pi-conjugated molecules known for their successful integration in optoelectronic devices.[15,16] The lateral order of the 2D blends and the molecule-substrate distances have been characterized using scanning tunneling microscopy (STM) and normal-incidence XSW, respectively. The former evidences highly crystalline molecular networks, in which each molecule is surrounded by the opposite species. The latter reveals how the molecules in the blends significantly change their molecule-substrate distances with respect to those in single component layers. Finally, we analyze the impact of those changes on the interface electronic properties by means of photoemission work function variation measurements.

EXPERIMENTAL DETAILS

All experiments were performed in ultrahigh vacuum chambers with a base pressure in the low $10^{-10}$

mbar range. The Ag(111) and Cu(111) substrates were cleaned by standard Ar-ion sputtering and annealing cycles. Evaporation of molecules onto the clean substrates held at room temperature was monitored by a quartz crystal microbalance.

STM measurements were performed at room temperature on Cu(111) and at 80 K on Ag(111), using a commercial Omicron VT-STM and electrochemically etched tungsten tips in constant current mode. STM images were analyzed using the WSxM software.[17] Errors in the determined unit cell parameters are the standard deviation obtained from several STM images.

Normal Incidence X-ray Standing Wave measurements were performed at the ID32 beamline of the European Synchrotron Radiation Facility in Grenoble, France. The beamline is equipped with a SPECS Phoibos 225 hemispherical electron analyzer, mounted at 90º with respect to the incoming beam, that can reach kinetic energies up to 15 keV and has energy resolution down to $\Delta E/E=10^{-6}$. The XSW measurements required photon energies around 2628.6 eV for Ag(111) and 2970.2 eV for Cu(111), corresponding to the (111) Bragg reflections of said substrates (111-layer spacings $d_0(Ag)=2.3586$ Å and $d_0(Cu)=2.0871$ Å). In order to minimize potential beam damage (see supporting information for more details), irradiation time was kept as low as possible during each measurement, and subsequent measurements were taken at different positions on the sample. The resulting reflectivity and photoelectron yield curves were fitted using the pyXSW software developed by Jerôme Roy. Due to the system's geometry (see above), multipole correction parameters are close to zero and may be disregarded.[11,18,19]

A commercial SPECS 10/35 UV source was used for the UPS measurements. The angle of incidence was 45º and electrons were detected close to normal emission. The sample was biased at -24.22V in order to shift the spectrum to higher kinetic energies to avoid instrumental modification of the analyzer transmission due to stray electric and magnetic fields. The cut-off position was determined by a sigmoid fit. The energy position of the cut-off in different measurements of the clean substrates was found to be reliable, giving a standard deviation of 0.01 eV for Ag(111) and 0.03 eV for Cu(111).

RESULTS AND DISCUSSION

Constant current STM images show that depositing PFP, CuPc or molecular blends on Ag(111) and Cu(111) leads to the formation of ordered molecular layers in which molecules adopt a flat-lying configuration. The influence of the substrate is seen from the discrete azimuthal domains related to the substrate symmetry and observed in all cases. The structure of PFP or CuPc monolayers on Ag(111) and Cu(111) have already been reported on in the literature.[9,20-24] We therefore place our focus on the mixed layers. When co-deposited (or deposited sequentially one after the other, which renders the same results) in an approximately 1:1 ratio, the two molecules form an ordered structure similar to that found for closely related systems,[14,25,26] in which molecules of one type surround themselves by the other type, maximizing donor-acceptor contact and -C-F···H-C- intermolecular interactions (Fig. 1). The experimental unit cell parameters are summarized in Table 1. These, in combination with the measured domain orientations, allow us to put forth rectangular, commensurate structures as tentative epitaxial models. These are depicted in Fig. 1c, and the corresponding parameters are included in Table 1.

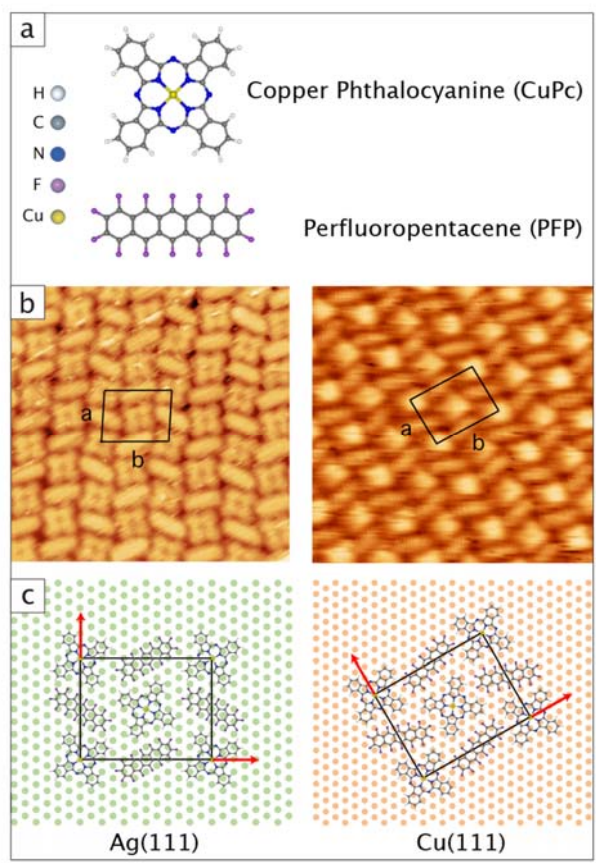

**Figure 1.** (a) Chemical structure of copper phthalocyanine (CuPc) and perfluoropentacene (PFP). (b) 11.5 nm x 11.5 nm images of the 1:1 molecular mixture on Ag(111) (left) and Cu(111) (right). Rectangular unit cells are marked in black. Scanning parameters -1.31 V, 0.010 nA on Ag(111) and -0.43 V, 0.034 nA on Cu(111). (c) Epitaxial relation between molecular overlayer and substrate. The high symmetry directions of the substrate (1-10) (close packed direction) and (11-2) are indicated by red arrows.

**Table 1**. Summary of STM results, including epitaxy matrices with base vectors (1 -1 0) and (1 1 -2).

|   | Ag(111) | | Cu(111) | |
|---|---|---|---|---|
|   | Experiment | Model | Experiment | Model |
| a | 22±2 Å | 23.1 Å | 21±1 Å | 22.2 Å |
| b | 29±1 Å | 30.0 Å | 27±2 Å | 28.2 Å |
| γ | 89±6° | 90° | 89±5° | 90° |
| Epit. Matrix |  | (8  0 / 0  6) |  | (0  5 / 11  0) |

Normal incidence x-ray standing wave measurements of the donor-acceptor mixtures were performed probing the C, N and F atoms. Upon X-ray irradiation of a single crystal in proximity to Bragg scattering conditions, the incident and scattered waves interfere to produce a standing wave with precisely defined periodicity (that of the scattering planes) and changeable phase. Atoms immersed in the standing waves show a photoemission yield that depends on their position relative to the wave field's nodal planes. XSW measurements consist in recording core-level photoemission yields in dependence of the standing wave phase, whereby one obtains element-specific information on the atomic location with respect to the substrate crystal planes.

Because of the molecular compositions (see Fig. 1a), the values extracted for the N and F atoms can be unambiguously ascribed to CuPc and PFP, respectively. However, a more complex scenario is found for C, which is contained in both molecules. Based on previous high-resolution XPS studies on PFP and CuPc monolayers and the 1:1 mixture,[14] the signal from each molecule should be disentangled as follows. Three separate peaks are resolved in the mixture's C1s photoemission spectra as illustrated in in Fig. 2 c). From highest to lowest binding energies these peaks correspond to PFP's C-F component (carbon atoms bound to fluorine), PFP's C-C component (carbon atoms bound solely to carbon) convolved with CuPc's C-N component (carbon atoms bound to nitrogen), and CuPc's remaining components, C-C and C-H (carbon atoms bound to solely to carbon or also hydrogen). In our analysis we do not consider the second, unfilled, peak, in which components of both molecules overlap, i.e. we only consider C-F in PFP (light grey filling) and C-C/H in CuPc (darker grey filling), which can each be fitted by a single peak and are therefore expected to give the most accurate results. The C-F atoms of PFP and C-C/H of CuPc are marked in grey in the molecular diagrams in Fig. 2c).

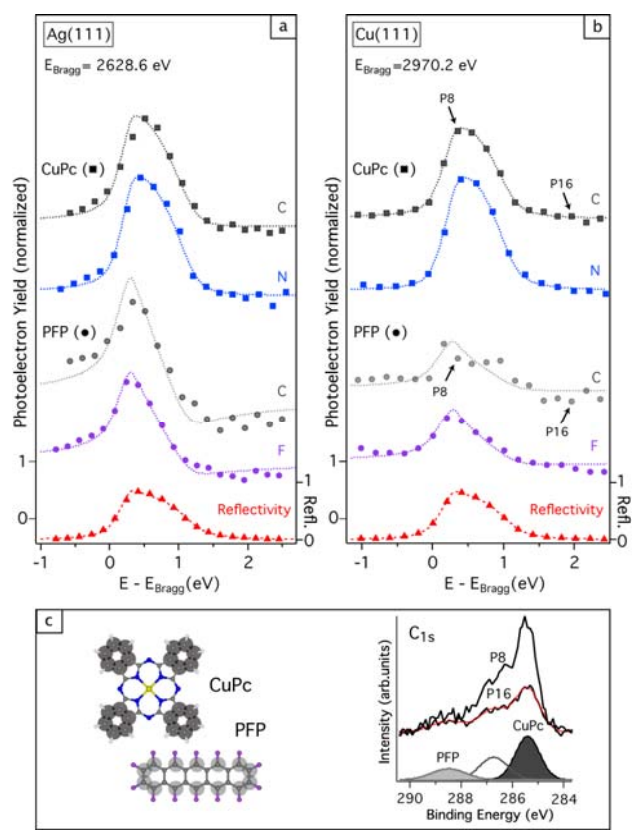

**Figure 2.** Reflectivity curve ("rocking curve", red triangles) and C1s, N1s and F1s photoelectron yield curves (grey, blue and purple curves; those of C and N are offset for clarity) for the molecular mixture on Ag(111) (a) and Cu(111) (b). (c) C1s photoemission intensity at the energies corresponding to P8 and P16 of the photoelectron yield curve [points indicated by arrows in (b)]. As shown for P16, the curves are fitted with three Gaussians, each corresponding to the different chemical environments of the carbon atoms in the molecules (see main text). The carbon atoms that were considered in the analysis are marked in grey in the molecular diagrams to the left (and correspondingly in the C1s spectrum to the right). These are CuPc's C-H and C-C components (dark grey) and PFP's C-F component (lighter grey). The remaining carbon atoms, PFP's C-C and CuPc's C-N, are convolved in the central Gaussian component and are not considered.

Fig. 2 shows an example of a reflectivity curve and of the photoelectron yield curves for CuPc's C-C/H (grey) and N (blue), as well as PFP's C-F (grey) and F (purple), obtained for each substrate.[27] The lower quality of the data corresponding to PFP's carbon (the smallest peak in the inset of Fig. 2) can be explained by a worse signal to noise ratio in the core-level spectra. From the fitting of the yield curves we obtain the coherent fraction (C.F.) and coherent position (C.P.). The C.F. is related to the degree of vertical order in the sample, i.e. the distribution of heights of each chemical species. If all the atoms of a given species had the same height above the substrate, the coherent fraction would be equal to one. Though coherent fractions very close to one can be measured in typical inorganic single crystals, organic layers are an entirely different matter, and coherent fractions under 0.5 are not unusual.[6-9] On the other hand, the C.P. is directly related to the adsorbate-substrate distance $d_H$, referred to the outmost surface plane:

$$d_H = d_0 \times (n + C.P.) \qquad (1)$$

where $n$ is a natural number and $d_0$ is the periodicity of the standing wave. As can be seen from the above expression, $d_H$ can only be determined within a multiple of $d_0$. In the case of organic molecular (sub)monolayers, expected adsorption heights are in the range between 2.2 and 3.6 Å.[28] $d_0$ is therefore sufficiently large to make all $n$ values but one unreasonable. Here we safely assume $n = 1$ in all cases.

C.F., C.P. and $d_H$ values for all analyzed species are summarized in Table 2 and a schematic representation of the molecular heights in the mix is shown in Fig. 3. The heights previously reported for single component CuPc and PFP layers [7-10] are also included for comparison. The molecule-substrate distance generally mirrors the interaction strength between the molecule and the underlying surface. This has been reported for instance for CuPc ,PTCDA or DIP, all of which show a clear trend of decreasing height on increasingly interactive substrates when going from Au(111), through Ag(111) to Cu(111).[9-12] In line with those observations, all our measured heights show consistently lower values on Cu(111) than on Ag(111). However, disregarding that difference, results on Ag(111) and Cu(111) are qualitatively similar: the most pronounced changes comparing single component [7-10] and mixed layers are found in the acceptor molecule PFP. Our analysis reveals a substantial height change in both the C-F and F atoms, suggesting that the entire molecule is raised ~0.3 Å from the surface in the mixture. On the other hand, the height of the CuPc atoms does not change substantially.[9,10] In the pure CuPc layer on Au, Cu and Ag,[9,10] N remains slightly lower than the C atoms. This is expected, as they form the central cage with the Cu atom that interacts most strongly with the metal substrate. In the mixture, this configuration, as well as the height of the CuPc molecules, remains practically unchanged, indicating a stronger interaction with the substrate than for PFP.

In single-component layers, CuPc lies closer to the substrate surface than PFP, both on Ag(111) and Cu(111).[7-10] Upon blend formation, the raising up of PFP further increases the height difference between donors and acceptors (Fig. 3). The intermolecular C-F⋯H-C interactions assumed to drive the self-assembly of the highly crystalline donor-acceptor networks would be strongest in a coplanar arrangement, which reduces the bond distance and enhances its linearity.[29] They are therefore expected to tend to level the molecular heights in the blends. However, contrary to expectations, we find that the height difference between molecules is increased in the mixed layer. The driving forces behind these surprising changes are unclear. Our hypotheses point either to substrate-mediated effects or to halogen-π interactions between the PFP fluorines with the π-orbitals above the CuPc (C-F⋯π). The former seems most intuitive and may arise from changes in the interface electronics,[14,30] which in turn modify the molecule-substrate interactions and the associated adsorption distances. The latter would profit from an intermolecular height offset in the order of that found experimentally. As opposed to the C-F···H-C interactions, which would tend to level the molecular heights and are known to be amongst the weakest hydrogen bonds,[29,31,32] C-F···π interactions have been shown to play an important role in organic crystal packing and their additional contribution to the intermolecular interactions therefore seems another plausible explanation to our findings.[31]

**Table 2**. Average coherent fraction (C.F.), coherent position (C.P.) and height above the substrate ($d_H$) of the different atomic species (see supplementary information for information on errors).

|  |  | C-F | F | C-C/H | N |
|---|---|---|---|---|---|
| Ag(111) | C.F. | 0.8 | 0.59±0.08 | 0.6 | 0.73±0.04 |
| | C.P. | 0.48 | 0.49±0.02 | 0.34 | 0.301±0.008 |
| | $d_H$ | 3.48±0.05 Å | 3.52±0.05 Å | 3.16±0.05 Å | 3.07±0.05 Å |
| Cu(111) | C.F. | 0.3±0.1 | 0.32±0.09 | 0.48±0.07 | 0.7±0.1 |
| | C.P. | 0.59±0.04 | 0.62±0.02 | 0.30±0.02 | 0.28±0.01 |
| | $d_H$ | 3.33±0.07 Å | 3.37±0.05 Å | 2.72±0.05 Å | 2.66±0.05 Å |

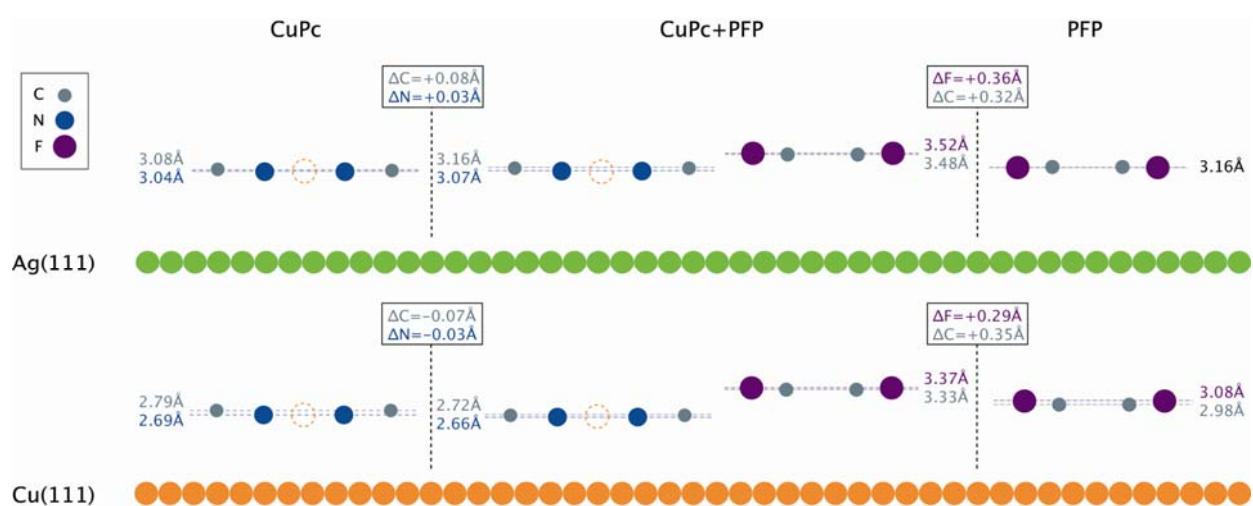

**Figure 3.** Schematic representation of molecule adsorption heights with respect to the Ag(111) (top) and Cu(111) (bottom) surfaces in the case of CuPc monolayers (left),[9,10] the PFP monolayers (right),[7,8] and the mixed donor-acceptor layer (center). Height changes in the mixed layers referred to the single component monolayers are included. The distance to the substrate is not to scale.

On the whole, the increased molecule-substrate distance of PFP and unchanged distance for CuPc suggest an overall reduced interaction of the mixed molecular layer with the underlying substrate. Reduced molecule-substrate interactions as a result of enhanced intermolecular interactions in molecular mixtures have been reported before and are consistent with general chemical concepts.[33,34] However, this is the first quantitative report on the associated changes in the molecule-substrate distances of donor-acceptor blends, which are in turn of utmost importance for the understanding of the interfacial electronic properties.[1,2,14]

The changes in molecular height and conformation found in the mixed layer are expected to lead to variations in the interface dipole. These variations are typically measurable as a shift in the system's vacuum level. We investigate this effect by systematic measurements of work function variations obtained from the secondary electron cut-off in ultraviolet photoemission (UPS) spectra. The change in the work function in dependence of molecular coverage was first determined for each single component of both substrates. The sample's work function was measured first for the clean substrate as reference, and then again after each subsequent stepwise evaporation. The results are shown in Fig. 4a. In all cases the work function is found to decrease steadily up to certain coverage, after which it remains practically constant. This "saturation" point coincides with the monolayer (ML) coverage, at which the surface is completely covered with molecules. The work function shift for 1ML on Ag(111) and Cu(111) was found to be 0.41 eV and 0.33 eV for PFP, in agreement with previous studies [7,8] and 0.44 eV and 0.72 eV for CuPc.

We now turn our attention to the work function of the mixed layers. Fig. 4b shows the vacuum level shifts associated with the deposition of the CuPc+PFP molecular mixture and of just PFP. The same quantity of PFP (namely approximately the one needed to form the binary network) on both the CuPc/Metal and the clean metal is found to shift the vacuum level by different amounts. On Cu(111), 0.42 ML of PFP was deposited onto 0.48 ML CuPc/Cu(111), producing a shift of $\Delta\phi_{mix}(Cu)=-0.10$ eV in the work function. From the known coverage dependence of the work function of the PFP/Cu(111) system, we find that $\Delta\phi_{mix}(Cu)$ is a *ca.* 40% reduction of the expected value for a direct 0.42 ML PFP deposition on Cu(111) $\Delta\phi_{pure}(Cu)=-0.17$ eV.[35] The scenario on Ag(111) is very similar: 0.32 ML of PFP deposited onto 0.48 ML CuPc/Ag(111) causes the work function to shift downwards by $\Delta\phi_{mix}(Ag)=-0.10$ eV, compared to an expected value of $\Delta\phi_{pure}(Ag)=-0.16$ eV. Again, a reduction of about 40% is found.

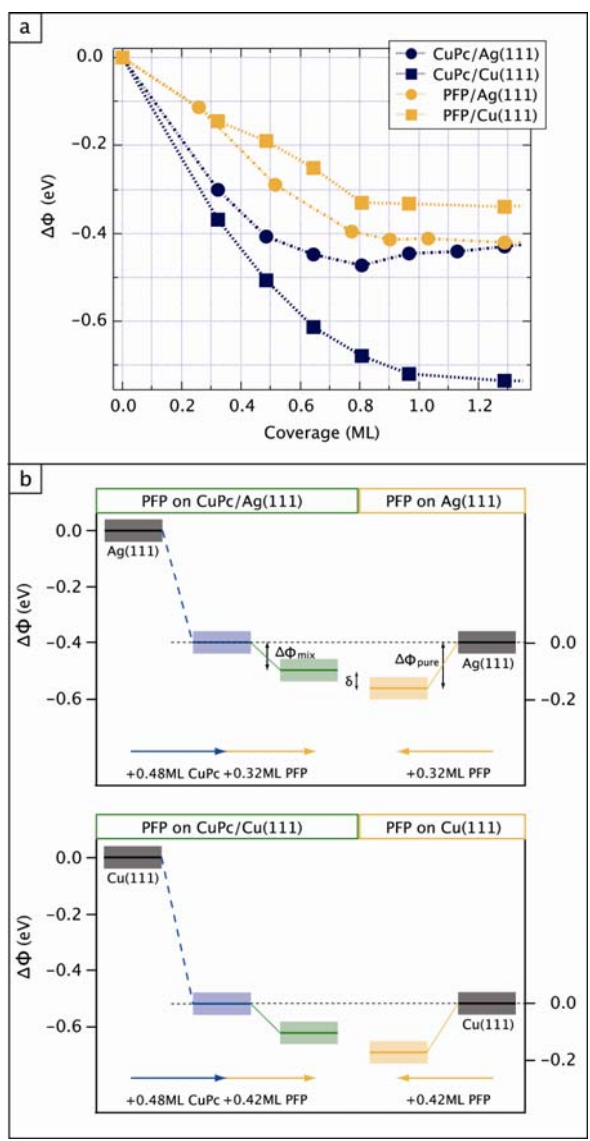

**Figure 4.** (a) Work function changes associated with CuPc (blue) and PFP (yellow) on Ag(111) (round markers) and Cu(111) (square markers) substrates as a function of coverage. (b) The work function of the clean metal is reduced upon deposition of molecules. Deposition of a certain amount of PFP onto the clean metal results in a shift $\Delta\phi_{pure}$. Depositing the same amount of PFP onto a pre-covered CuPc/Metal system (with adequate CuPc coverage to render the formation of submonolayers of a 1:1 stoichiometric blend) results in a further shift of $\Delta\phi_{mix}$. Experimentally, we find a difference $\delta$ between $\Delta\phi_{pure}$ and $\Delta\phi_{mix}$ ($\Delta\phi_{pure} > \Delta\phi_{mix}$). The blue bar on the left (CuPc/Metal) and the black bar on the right (clean metal) have been aligned in order to more easily compare $\Delta\phi_{mix}$ and $\Delta\phi_{pure}$. The error in $\Delta\phi$ is represented by the shaded area.

The reduction should not be attributed to the deviations from linearity of the work function variations at coverages close to or above the monolayer. A statistical analysis of 14 STM images of the CuPc+PFP/Ag(111) sample revealed 72% of the surface to be covered with the 1:1 mixture, while the remaining area appeared to be empty (26%) or covered with patches of ordered PFP (2%), i.e. total coverage reached only 74% of a ML.[36] At this coverage, the non-linearity of the work function dependence cannot be responsible for the changes in vacuum level.

Instead, this effect should be understood in terms of differences in the interface dipole due to the different geometries of the molecules in the mixed and single component layers. As demonstrated by XSW measurements, geometry changes only affect the PFP molecule, and hence we may assume that only the PFP dipole changes when going from pure layers to the mixture. The change in the effective PFP dipole may be obtained from the experimentally determined work function changes $\Delta\phi_{pure}$ and $\Delta\phi_{mix}$ by using the Helmholtz equation [7]

$$\Delta\phi = e P n / \varepsilon_0 \qquad (2)$$

where P is the effective dipole moment per molecule, $n$ is the areal density of dipoles,[37] $e$ is the elementary charge and $\varepsilon_0$ is the vacuum permittivity, we estimate that upon mixing, the effective dipole moment associated with PFP changes from $P_{pure}$=1.75 D to $P_{mix}$=1.10 D on Ag(111) and from $P_{pure}$=1.52 D to $P_{mix}$=0.90 D on Cu(111).

The most important effects that contribute to the interface dipole are (1) the Pauli repulsion between molecule's orbitals and the metal's electrons decaying into vacuum, (2) charge transfer between molecule and substrate and (3) the molecules' intrinsic electric dipole moment.[38] We discard option (2) as a possible explanation for the observed changes, since charge transfer values calculated on Ag(111)[39] indicate that this effect results in a net dipole change in the opposite direction. Option (3) is likewise ruled out as a main contributor to the reduction in P, since changes in the intrinsic dipolar moment in the mix due to modified molecular distortions lead to net dipole changes in opposite directions on Ag(111) and on Cu(111) [on Ag(111),[7,8] when going from single component to mixed layer, the net change in intramolecular dipole points into the surface, whereas on Cu(111) it points away from it]. We therefore argue that it is mostly the Pauli repulsion (1) that is responsible for the observed changes in P: the increased molecule-substrate distance of PFP found in the mixture translates into a reduced Pauli repulsion, decreasing the effective interface dipole, as measured experimentally.

CONCLUSIONS

In conclusion, we have characterized the lateral and vertical structure of CuPc and PFP molecular mixtures in a 1:1 ratio on Ag(111) and Cu(111) substrates. Contrary to what might be expected in a molecular network stabilized by hydrogen bonding, the first XSW measurements on multicomponent systems reveal that CuPc and PFP lie at considerably different heights. Most strikingly, that difference is enhanced in the mixed layers as compared to the respective heights in single component layers. While CuPc remained virtually at the same height upon mixing, PFP was found to raise up substantially (ca. 0.3 Å) on both the Ag(111) and Cu(111) substrates. Such a change in the adsorption height of PFP is expected to affect interface phenomena. Our photoemission measurements show this effect is indeed measureable, as we find that the work function shift caused by deposition of PFP onto CuPc/Metal is smaller than what is found for the deposition of the same amount of PFP onto the bare metal substrate. We hereby provide a direct measure of the effect of a molecule's adsorption height on vacuum level shifts and, in turn, interfacial energy level alignment.

ASSOCIATED CONTENT

Supporting information: XSW error analysis and beam damage (Figures S1), work function error analysis. This material is available free of charge *via* the Internet at http://pubs.acs.org.

ACKNOWLEDGEMENTS

This work was supported by the Spanish MICINN (MAT2010-21156-C03-01 and -C03-03), the Basque Government (IT-621-13), and Spanish Grant PIB2010US- 00652. D. G. O. acknowledges support from the European Union under FP7-PEOPLE-2010-IOF-271909. We thank the European Synchrotron Radiation Facility (ESRF) for provision of synchrotron radiation facilities, J. Zegenhagen, H. Isern and L. Andre for support of the XSW beamline, and Jerôme Roy for his help with the pyXSW data analysis software. E. G. thanks J. G. Goiri for technical assistance.

# Table of Contents (TOC)

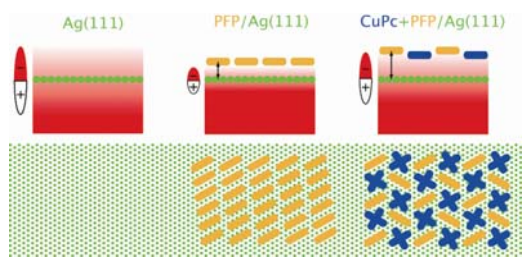